\documentclass[preprint,12pt,3p,authoryea]{elsarticle}
\setcitestyle{authoryear,open={(},close={)}}

\usepackage[title]{appendix} 

\usepackage{natbib}

\usepackage{amssymb}
\usepackage{subfig}
\usepackage{color}
\usepackage{hyperref}

\begin{document} 

\begin{frontmatter}

\title{\large{The Power of Trading Polarity: Evidence from China Stock Market Crash}}

\author[label1,label2]{Shan Lu}

\author[label1,label3]{Jichang Zhao\corref{cor1}} 
\ead{jichang@buaa.edu.cn}

\author[label1,label3]{Huiwen Wang}

\address[label1]{School of Economics and Management, Beihang University, Beijing, China}
\address[label2]{Beijing Key Laboratory of Emergency Support Simulation Technologies for City Operations, Beijing, China}
\address[label3]{Beijing Advanced Innovation Center for Big Data and Brain Computing, Beijing, China}
\cortext[cor1]{Corresponding author}

\begin{abstract}
The imbalance of buying and selling functions profoundly in the formation of market trends, however, a fine-granularity investigation of the imbalance is still missing. This paper investigates a unique transaction dataset that enables us to inspect the imbalance of buying and selling on the man-times level at high frequency, what we call `trading polarity', for a large cross-section of stocks from Shenzhen Stock Exchange. The trading polarity measures the market sentiment toward stocks from a view of very essence of trading desire. When using the polarity to examine market crash, we find that trading polarity successfully reflects the changing of market-level behavior in terms of its flipping times, depth, and length. We further investigate the relationship between polarity and return. At market-level, trading polarity is negatively correlated with returns, while at stock-level, this correlation changes according to market conditions, which becomes a good signal of market psychology transition. Also, the significant correlation disclosed by the market polarity and market emotion implies that our presented polarity, which essentially calculated in the context of high-frequency trading data, can real-timely reflect the sentiment of the market. The trading polarity indeed provides a new way to understand and foresee the market behavior.

\end{abstract}

\begin{keyword}
stock market crash \sep trading behavior \sep trading imbalance \sep trading polarity \sep econophysics
\end{keyword}

\end{frontmatter}

\section{Introduction}
The 2015 China stock market crash had witnessed a-third of the A-shares market value losses within one month after huge amounts of panic sell-off. It again demonstrates the non-negligible roles of the trading behavior, especially when inexperienced investors dominate the market \cite{zhou2017}. The stock market crash, usually triggered by economic events, is thereafter is led by crowd behavior and psychology such as mimicking trading fashion \cite{shiller1987, lu2017herding,zhao2011herd}, severe overlapping of portfolios \cite{delpini2018network} and etc.. Among these trading behaviors, one of the most important aspects behind market microstructure is the imbalance of selling and buying parties. 

An intuitive measurement of imbalance, called herding, is firstly proposed to study the trading behavior of institutions \cite{lakonishok1992}. It accounts for the inequality between the number of managers who cut their holdings and the number of those who increase holdings at the level of individual stocks with quarterly frequency. This measurement is widely used in the subsequent studies \cite{wermers1999,sias2004}. Not only in studies of institutional trading behavior, when examining individual investors trading pattern, similar measurement is also applied to measure marginal difference in the extent to which an investor's sales of a stock tend to parrot other individual investors' tendencies to sell the stock in the subsequent trading days~\cite{grinblatt2012iq}. A similar definition, named individual investor imbalance, quantifies the net individual trading effect and it has been proved to have predictive power with respect to abnormal returns on and after dividend announcements \cite{kaniel2008,kaniel2012}. The existing findings suggest that imbalance of selling and buying can be indicative or even predictive in understanding market trends. 

Nevertheless, the above two branches of measuring imbalance need the identity of investors, which is usually not available in common stock transaction datasets for the protection of privacy. Moreover, they only investigate one kind of investors, either institutional or individual. From the view of whether the seller initiates a trade or the opposite way, a measurement called order imbalance is defined further to evaluate the inequality of the market control power either by selling or buying party \cite{chordia2004order}. Differently but interestingly, they have demonstrated that when buyer-initiated order overwhelmed the seller-initiated order in the previous trading day, the higher the return is today. It implies that instead of exploring the imbalance at the investor level, drilling down by probing the imbalance at the granularity of order might suggest a new manner of seeing the big picture of stocks as well as market with rich details reserved. 

Motivated by the above studies, we derive an indicator called trading polarity by taking the advantage of the transaction-level data in Shenzhen Stock Exchange. The trading polarity indicator is defined as the difference of buying and selling man-times, rescaled by the total man-times involved in a given time interval. Unlike the imbalance indicators defined in the previous studies that usually only reflect the imbalance of one kind of investors and at best on daily basis, the presented one particularly reflects the degree of imbalance in selling and buying from a man-times view and particularly in high frequency. It is worth noting that the man-times of transactions is the most micro unit to depict trading behavior. It captures the minimum decision unit of market participants without leaking the traders' identity. Moreover, as we measure the trading polarity on a one-minute basis, the indicator provides not only cross sectional but also longitudinal detailed information on the imbalance phenomenon, which is pretty useful under the current circumstance of the wide range of high frequency trading and financial big-data \cite{dufour2000time,preis2011switching,xie2016quantifying,bhattacharya2017superstitious,menkveld2017shades,lillo2003econophysics,ivanov2014impact}. In a word, the defined indicator offers a brand new angle of the profound personality of a market. 

As trading polarity can signal excessive investors interest in stocks, thus it could be related to future returns and provide additional power beyond trading activity measures in explaining stock market. While many studies have investigated the imbalance without the crisis period \cite{chordia2004order,grinblatt2012iq,kaniel2008,kaniel2012}, we will specifically work out the strength and influence of man-times imbalance during stock market crash. What we are trying to find is the vibratory rates of trading polarity, from which the price oscillatory originates. To make it more specific, we are trying to answer the questions below:

1) what is the relationship between trading polarity and return in a systematic view? 

2) how does trading polarity reflect the phase-changing of market? 

3) is there an inner law of trading patterns emerges from behind the market in crash? 

To answer these, we use concepts from econophysics, since studies in econophycis in recent decades have shown their advantages in understanding the global behavior of economic systems without the preparation of a detailed microscopic description of the same system \cite{mantegna1999introduction,fan2004network,ivanov2014impact,wu2011socioeconomic,huang2015experimental}. We take up the statistical concepts, such as power-law distribution and burstiness, to investigate the stock market system at different scales. We find that at market-level, trading polarity is negatively correlated with return. At stock-level, this correlation changes day-to-day, according to market conditions. When using the polarity to examine market crash, we find that trading polarity successfully illustrates the transitions of market conditions in terms of its flipping times and depth. And the length before polarity flipping follows roughly stable power-law distribution, which exhibits the underlying rule behind trading activity. Moreover, the observed bursty character of length before polarity flipping reflects the potentially generic feature of trading dynamics at the time of government bailouts. Even more inspiring, the significant correlation disclosed by the market polarity and market emotion implies that our presented polarity, which essentially calculated in the context of high-frequency trading data, can real-timely reflect the sentiment of the market. Given the above, we argue that the trading polarity provides a new way to understand and foresee the market behavior.

The rest of the paper is organized as follows. Section \ref{subsec:polarity} puts forward the definition of trading polarity. Section \ref{subsec:data} gives detailed descriptions on transaction data and price data used in the study. Based on the definition and the unique transaction dataset, section \ref{subsec:statistics} presents the statistical properties of trading polarity from an econophysical view, where several important indicators are proposed to examine market crash. We then jump out of the polarity itself, to explore its effects on other market indicators at the market-level in section \ref{subsec:marketlevel} and at the stock-level in section \ref{subsec:stocklevel}. Finally, section \ref{subsec:conclusion} draws the conclusions.

\section{Trading polarity}\label{subsec:polarity}

When transactions are partially filled, there would be imbalance. For a given stock $i$ in time interval $[t-1, t]$, we define the trading polarity as

\begin{equation}
\textrm{polarity}_{i,t} = \frac{\textrm{number of buying man-times}_{i,[t-1,t]}-\textrm{number of selling man-times}_{i,[t-1,t]}}{\textrm{number of buying man-times}_{i,[t-1,t]}+\textrm{number of selling man-times}_{i,[t-1,t]}}.
\end{equation}

This indicator can reflect the imbalance of selling and buying man-times in a specific time interval during a trading day. Being scaled by the total number of buying and selling man-times of stock $i$ in $[t-1,t]$ makes it possible to compare the extent of imbalance among different time intervals, and among different trading days, regardless of the market environment. It is also comparable among stocks, as the indicator ranges from -1 to 1. A positive value of polarity reveals that buying man-times overwhelms the selling man-times. More specifically, a positive polarity denotes an imbalance between buyers and sellers in which buying man-times exceeds selling man-times, which we call it `buying polarity'. There is a net buying man-times flow. A negative polarity, called `selling polarity', is the imbalance in which selling man-times exceeds buying man-times. There is a net selling man-times flow. Considering the fact that in realistic dataset, only transactions that had already been done are recored, the imbalance on the number of trading parties could reflect: (1) given a certain amount of trading volume, it is revealed by the sign of polarity that whether the trading is concentrated in the hands of a fewer sellers than buyers or is the other way around; (2) how concentrated of the transactions, or say, the extent of imbalance, revealed by the absolute value of polarity. 

Broadly speaking, the trading polarity indicates which party is more crowded or potentially have more investors that agree on whether it is time to sell or to buy. For instance, the indicator can be high either due to a wide range of investors are buying or a small number of investors are selling the stock. Thus it could measure how market participants anticipate the price trend as well as the outcome of multi-player gaming of trading under complex circumstances within the specific time interval. In another word, the net number of trading man-times (includes all market participants, such as institutional and individual investors) indicates the market sentiment towards stocks. 

In contrast to those definitions which regard sellers or buyers by specific different kinds of traders, we argue that the number of buying man-times and the number of selling man-times represent the very essence of trading desire in an extremely short time, for example, in one minute in particular. That is, we treat the man-times as the most micro decision unit of transaction no matter they are conducted by one or couples of people, which is crucial in explaining the imbalance of trading parties. 

\section{Data}\label{subsec:data}

\subsection{Sample period: 2015 stock market crash of China}

\begin{figure}[h]
\centering
\includegraphics[width=1\linewidth]{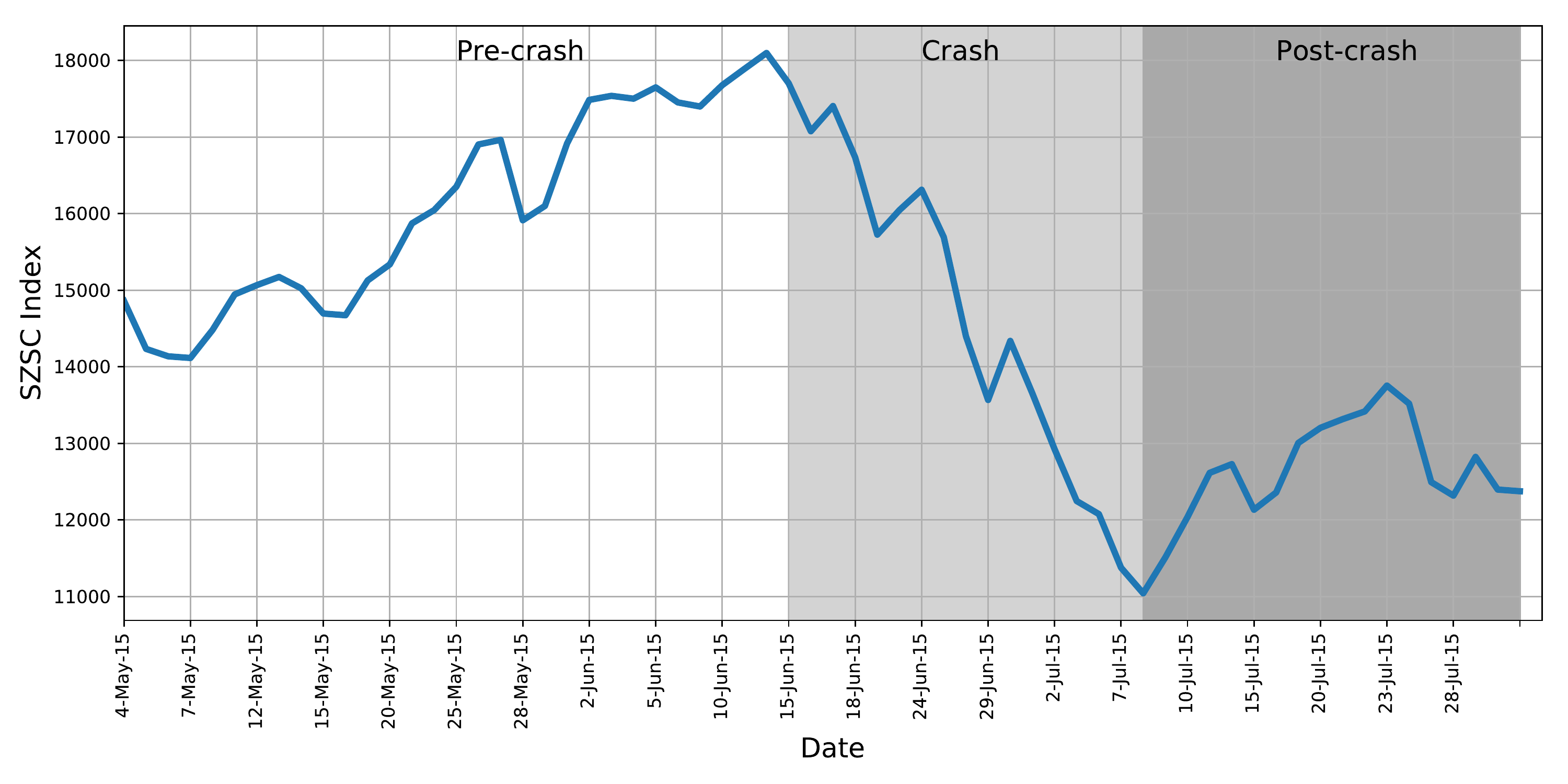}
\caption{{\bf The Shenzhen Stock Exchange Component (SZSC) index from May 4 to July 31 in 2015.} The SZSC index is an index of 500 stocks that are traded at the Shenzhen Stock Exchange. It is one of the main stock market indices in China stock market. The index shows that the market experienced a big rise and fall in this segment. As of June 12, there was a peak at 18098.27 points. From June 15 to July 6, this figure experienced a sharp shrank. Thereafter, the index reached its lowest in July and rose a little during July 8 to July 31. We cut the sample period into three parts in this order, shown by different backgrounds.}\label{fig:market_index}
\end{figure}

It has shown that the systemic risk is higher during crashes in 2001 and 2008 than in calm periods for China stock market \citep{ren2014dynamic}. The 2015 China stock market crash is generally considered the worst stock market crash since 2007. As we mainly focus on the behavior around market crash, we narrow down the sample period between May 4 2015 and July 31 2015. As can be seen in Fig.~\ref{fig:market_index}, the stock market experienced a big rise and fall in this segment as revealed by the Shenzhen Stock Exchange Composite(SZSC) index. At the beginning, people trusted the market and this coincided with a subsequent rise in prices. It is revealed by the SZSC index that May 5 to June 12 had witnessed an unprecedented bull market as the index kept rising up to the top, what we call `pre-crash' period. 

As the stocks became over-priced, the bull market came to an end and the beginning of the crash started from June 15 2015. Market participants started from not believing the market is going down to abrupt panic sell-off, triggering sharp falling of prices within a sudden and panic again followed. Roughly speaking, the first throes of the crash ended around July 7 2015. We name it as `crash' period. 

The second throes of the crash ranges from July 8 to July 31, in which government had enacted many measures such as limiting short selling, making mutual funds pledge to buy more stocks \citep{france2015almost}. However, those tremendous orchestra were of little success, revealing by only a little rise in stock index as shown in Fig.~\ref{fig:market_index}. We name it the `post-crash' period. 

The three-month period contains both bull market and bear market, which is perfect for the present study. In the following analysis, the sample is divided into three parts in the chronological order as described in the above. 

\subsection{Transactions records}

The data employed in the study contain transactions happened in Shenzhen Stock Exchange between May 4 2015 and July 31 2015, covering 1,471,848,085 records of transactions and 1646 stocks. The dataset consists of the stock ID, price, number of shares traded, time for every trade, and most importantly, the serial numbers for selling and buying orders. Unlike the previous studies, the unique of our data is that, for each selling or buying order (a one-time order involves price and volume for one stock), an identical serial number will be assigned to it according to the quote sequence. When the quote order being fulfilled or partially fulfilled, the certain serial numbers of the buying and selling orders are recorded in the transaction record. If the order is partially filled, the serial number could appear multiple times in the transaction records, as it is fulfilled by several counterparties' orders. Consequently, the serial numbers are reset at the beginning of trading days, and they rely solely on the order time, regardless of stocks. The serial numbers make it possible for us to distinguish potentially different trading decision units and count the number of man-times for buying and selling, given the transaction records in a specific time interval. Otherwise, one may argue that `for every buyer, there's a seller'. 

Admittedly, it is not the first time for transactional data employed in financial studies. Most of the datasets in these studies consist of stock ID, trading price and volume, and also the timestamp of trading \citep{wood1985investigation,dufour2000time,huang2003trading,lillo2003econophysics,xie2016quantifying}, as we do. However, to our best knowledge, it is the first time for man-times selling and buying kind of data to be studied. There are at least two advantages of using man-times information to capture the imbalance between buying and selling. First, the serial numbers of buyers and seller could be available in transactional data not only in China but also in other countries without leaking the privacy of traders. As far as we know, there are limited number of studies in stock market field contain the traders' information in their dataset \citep{grinblatt2001distance,grinblatt2009sensation,grinblatt2012iq}. In fact, not all countries, including China, allows the data with traders ID involved to go public, even within a small range of financial companies. Using man-times makes it easier for the imbalance indicator to be applied in practice, especially in course of the big-data era. Second, using man-times provides a more systematic view to measure the imbalance on the whole. In other words, instead of emphasizing trader-level individual effects on transaction activity, the man-times-level of trading polarity treats every trading decision unit equally and integrates them together to offer a global perspective on trading activities. Though the trading polarity origins from `micro' data, it could reflect the `macro' landscape of stock market imbalance. In section \ref{subsec:statistics} and \ref{subsec:marketlevel}, we show the usefulness of the proposed indicator in explaining the whole market picture.

To be more specific, we compute the polarity for each stock at one minute frequency. For simplicity, only the consecutive trading hours are included in the sample, ranging from 9:30am to 11:30am and from 13:00pm to 14:57pm, consisting of 237 minutes per day. Thus, we have the polarity time series for 1646 stocks listed on Shenzhen Stock Exchange for analysis (the sample has excluded stocks that have no trading at all during the sample period). We argue and show in the rest of the paper that the proposed indicator is an effective and practical tool with which to observe market microstructure in high frequency data.

\subsection{Stock prices}

The data of stock prices are downloaded from Thomson Reuters' Tick History. Two types of price time series data are used, including \emph{end-of-day} and \emph{intraday}. On one hand, we use the closing price from the \emph{end-of-day} data. It is the baseline price for computing the daily return for the next day, denoted as $p_{i,d}$. The daily percentage change of stock $i$ on day $d$ is computed by $(p_{i,d}-p_{i,d-1})/p_{i,d-1}$. The reason is that this kind of percentage change is consistent with what investors see during trading days on any trading information board, which could stir up tensions and impact the prices of stocks through trading behavior directly. This measure is used in section \ref{subsec:flippingtimes}. On the other hand, we use the last price of every one minute from the \emph{intraday} data in order to calculate the intraday percentage changes of stocks. The price of stock $i$ at time $t$ on day $d$ is $p_{i,t,d}$. The log-return of stock $i$ at time $t$ on day $d$ is $r_{i,t,d}=log(p_{i,t,d})-log(p_{i,t-1,d})$, as commonly used in most financial studies. This return measure is applied in section \ref{subsec:stocklevel} for analysis on stock-level polarity.

The datasets analyzed during the current study are publicly available in the figshare.com repository, and can be accessed freely through https://doi.org/10.6084/m9.figshare.5835936.v1.

\section{Statistical properties of trading polarity}\label{subsec:statistics}

Investigating the statistical properties of trading polarity and its variation is crucial for understanding the underlying mechanism behind the complexity of trading behavior. This section devotes to the properties of trading polarity and developing its extensions.

\subsection{The dominant direction of trading polarity}
\begin{figure}[h]
\centering
\includegraphics[width=0.6\linewidth]{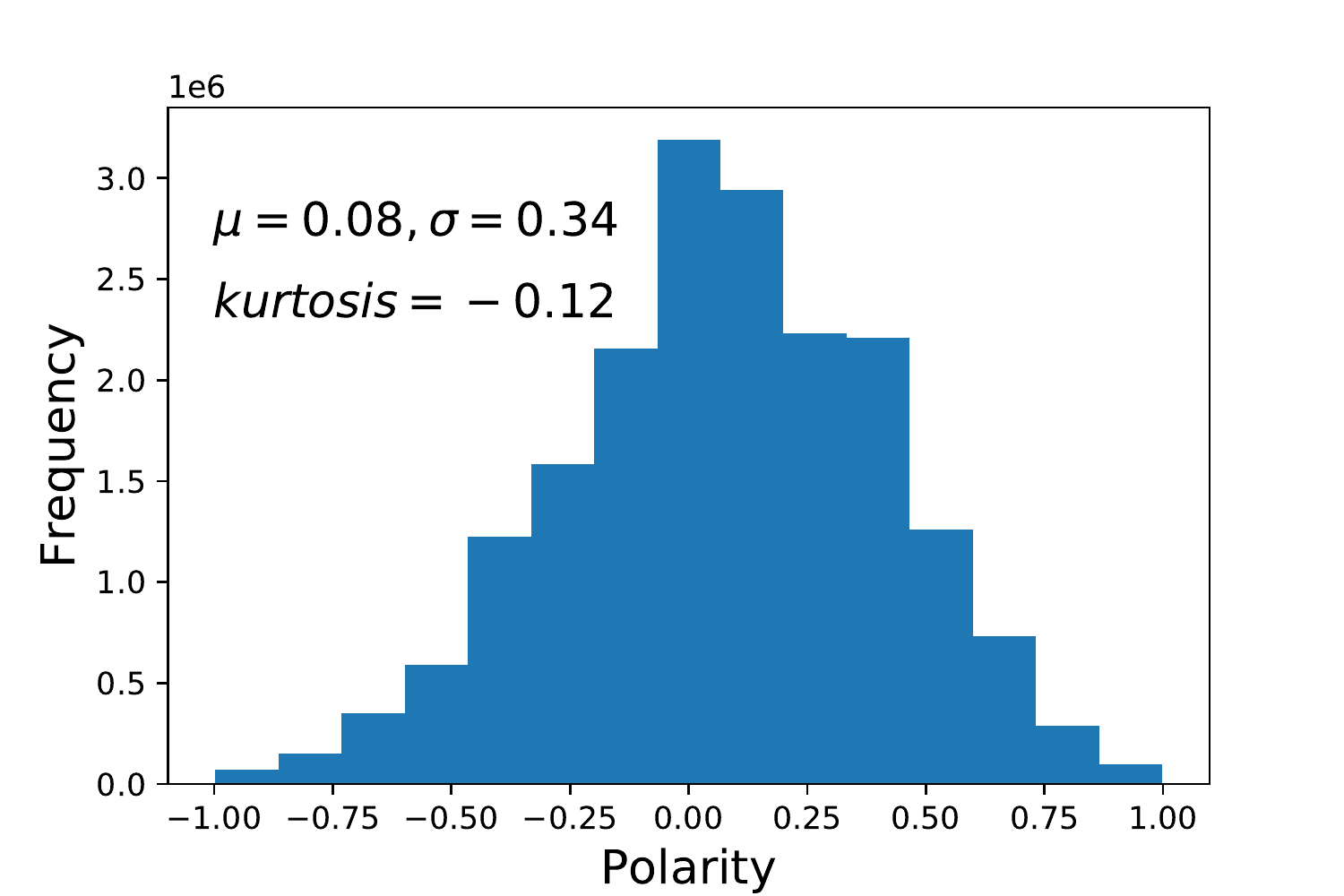}
\caption{{\bf The distribution of trading polarity.} It contains all the one-minute polarities in Shenzhen Stock Exchange between May 4 and July 31 2015. The mean ($\mu$), standard deviation ($\sigma$), kurtosis are given in the figure.}\label{fig:imbalance_distribution}
\end{figure}

The time flow of the trading polarity is far from uniform. To get a glimpse of the whole picture, we presents the distribution of all the one-minute polarities in Shenzhen Stock Exchange between May and July 2015 in Fig.~\ref{fig:imbalance_distribution}. The mean of polarity is 0.08, suggesting that market favor in buying man-times overwhelmed selling man-times on average. Great variations and platykurtic of the distribution can be found in standard deviation $\sigma=0.34$ and tailedness $kurtosis=-0.12$, which gives rise to the probability of polarity being in high-variability. One could expect that the stocks price movements are composed of their polarity oscillatory, as investors changing their strategy \citep{mizuno2004traders}. From this oscillatory polarization in trading, a market structure will arise. We explore the related topics in section \ref{subsec:flipping}

\begin{figure}[h!]
\centering
\includegraphics[width=0.95\linewidth]{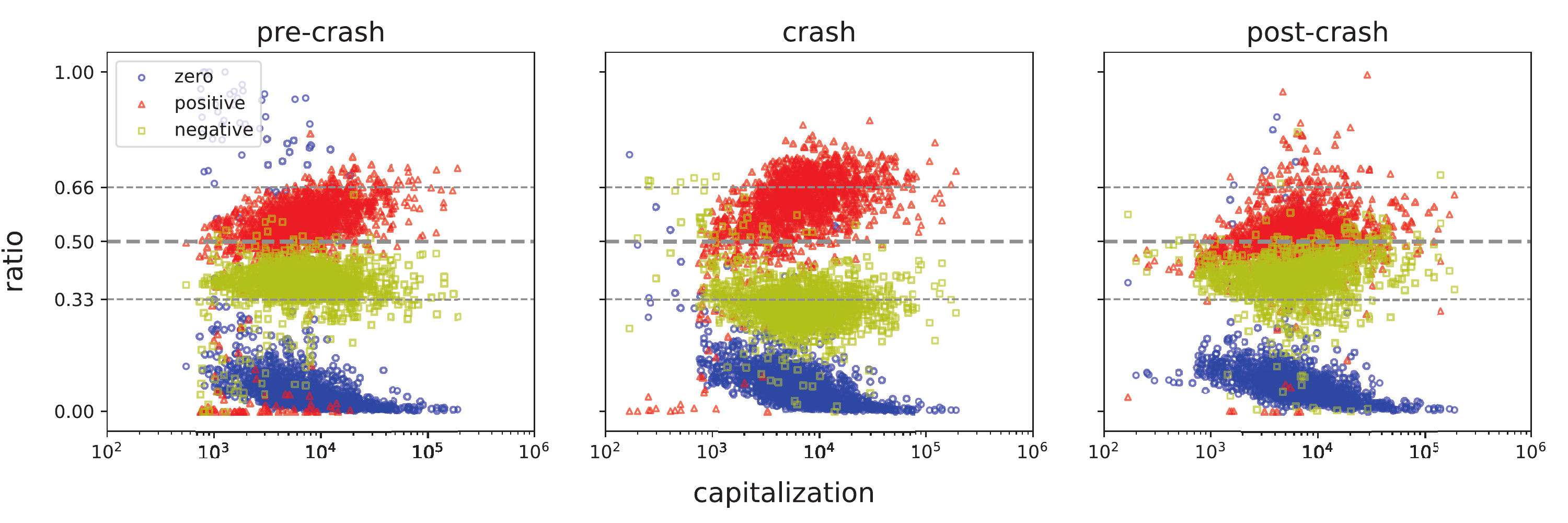}
\caption{{\bf Polarity direction ratios and stock capitalization.} We calculate the proportions of positive, negative, zero polarities for each stock in the certain period and plot the ratios in different colors on the y-axis along with their capitalization on the x-axis. Each group of polarities contains 1646 points representing all the stocks listed on Shenzhen Stock Exchange. Thus, every stock has three ratios in one subplot, and the three ratios add up to 1.}
\label{fig:capitalization_period}
\end{figure}

Basically, the symmetrical distribution of trading polarity is found in Fig.~\ref{fig:imbalance_distribution}. But does the symmetry hold for different stocks? To answer it, we calculate the proportions of positive, negative, zero polarities for each stock in the certain period. A higher ratio means the dominant role of the direction of polarity. The sum of the three ratios for one stock is assured to be 1. By dividing the sample period into three parts. Fig.~\ref{fig:capitalization_period} illustrates the capitalization of stocks against their polarity ratios. We have to use a logarithmic scale to display capitalization. It could be firstly noticed that the zero polarity have smallest proportions. This means that the China stock market is full of trading polarity, at least in the observation time window. Also, the positive ratios are over negative ratios across the three periods on the whole, indicating the probability of buying man-times overwhelmed the selling orders is larger than the opposite way, whatever the capitalization the stocks is and the market condition it is facing. Unlike positive and zero ratios, however, the negative ratios are invariant among stocks, indicating the selling polarity is quite much the same whatever cap the stocks are. 

As demonstrated in Fig.~\ref{fig:capitalization_period}, stocks with larger capitalization tend to have higher ratios on positive polarity rather than negative polarity, and the small ratio of zero polarity implies that it is not common for larger-cap stocks to have the same number of man-times orders from the two parties. In contrast, for the small-cap stocks, the difference between positive ratio and negative ratio is quite small. And it is relatively easier for small-cap stocks to have equal number of man-times orders on the two trading parties than for the large-cap stocks.

From the temporal view, the first and second periods both give evidence for that large-cap stocks are more correlated with polarity ratio than are other stocks. Especially in the second period, the difference between the positive ratio and negative ratio are largest, indicating a bit more dominant role the buying polarity plays at the beginning of market crash. In addition, the divergence among stocks with respect to the three kinds of ratios are bigger during market crash, which tells the story of the variations of trading polarity. The post-crisis period is rather different, whereas the positive and negative ratios both get closer to 0.5. This means that there's either a polarity of buying or selling rather than the balanced trading in post-crash period. The transitions between the two states present results on vibrations in collective trading pattern, which originates from the changing of market anticipation. 

Given these exploratory analysis, we can interpret the polarity indicator as a measure of market trading pattern, or furthermore, a measure of market-level irrational behavior. To deepen the understanding of the indicator, the following sections investigate not only the essence of its flipping roles but also its conjunctions with return and investors' emotion.

\subsection{Direction flipping of polarity}\label{subsec:flipping}

We believe that the trading polarity oscillations or vibrations will exhibit a rhythm. This rhythm configures the different market patterns that produced by the interplay between buyers and sellers. From the previous analysis, one could expect that the critical polarity levels constantly switch their signs among positive, negative, and zero. For instance, a former selling polarity, once it has been flipped, becomes a subsequent buying polarity in a subsequent downtrend; and an balanced polarity, once it has been penetrated, becomes either selling polarity or buying polarity in a later phase. The flipping cycle can vary in times, depth, length, from micro-oscillations to bull or bear market.

\subsubsection{Flipping times}\label{subsec:flippingtimes}

The first to discuss is the flipping times of polarity directions. Measuring them will enable us to recognize repetitive changes. The anomaly of these repetitive changes will give us clues to the unusual behavior of a market. 

For simplicity reasons, zero polarities are deleted in the original polarity series, especially considering their trivial occupations (see Fig.~\ref{fig:capitalization_period}). The flipping times for stock $i$ in day $d$ is obtained by counting how many times the polarity changes its sign. As different stocks are of different illiquidity and some of them may have no transaction at all in particular minutes, the polarity values are thus blanking in those minutes. Therefore, to ensure comparability, the flipping times are scaled by length of non-empty polarity values in a one-day series such that it only reflect the standardized flipping times, i.e.,

\begin{equation}
\textrm{standardized flipping times}_{i,d} = \frac{\textrm{number of flipping times}_{i,d}}{\textrm{length of effective polarity values in day }d}.
\end{equation}

\begin{figure}[h!]
	\begin{minipage}{0.33\linewidth}
		\centering
		{\footnotesize (a)pre-crash}
		\includegraphics[width=1\linewidth]{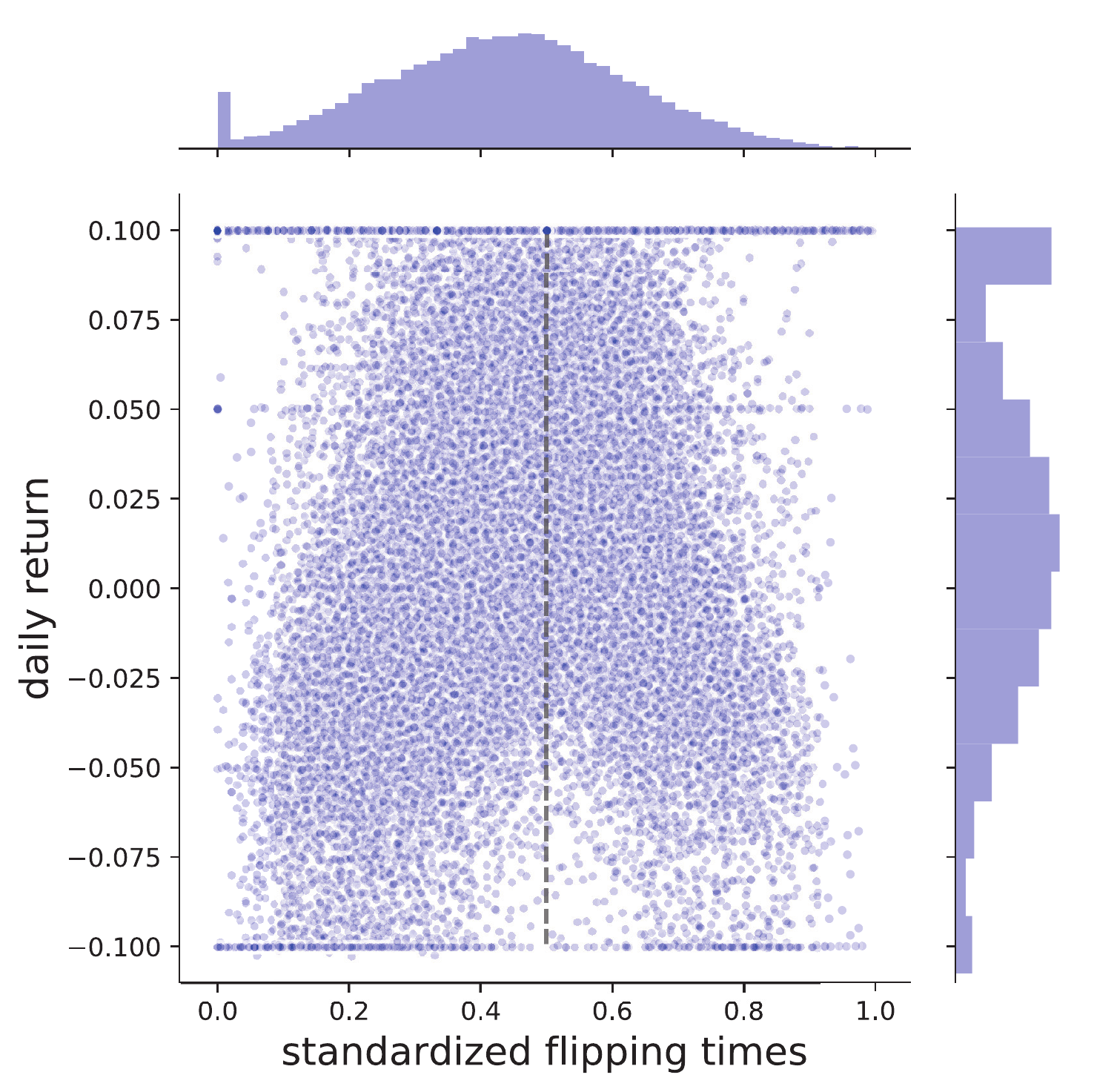}
	\end{minipage}\hfill
	\begin{minipage}{0.33\linewidth}
		\centering
		{\footnotesize (b)crash}
		\includegraphics[width=1\linewidth]{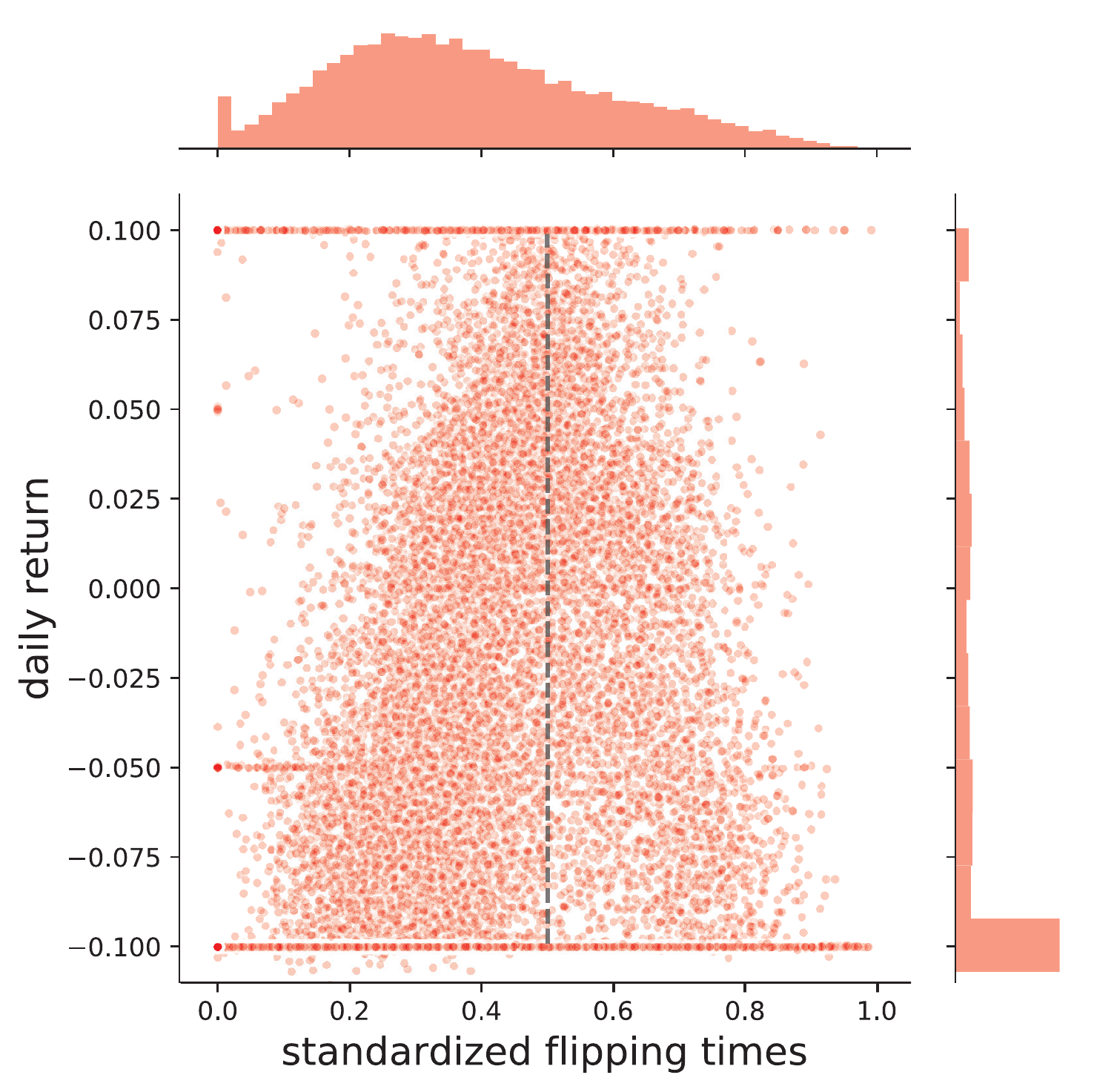}
	\end{minipage}\hfill
	\begin{minipage}{0.33\linewidth}
		\centering
		{\footnotesize (c)post-crash}
		\includegraphics[width=1\linewidth]{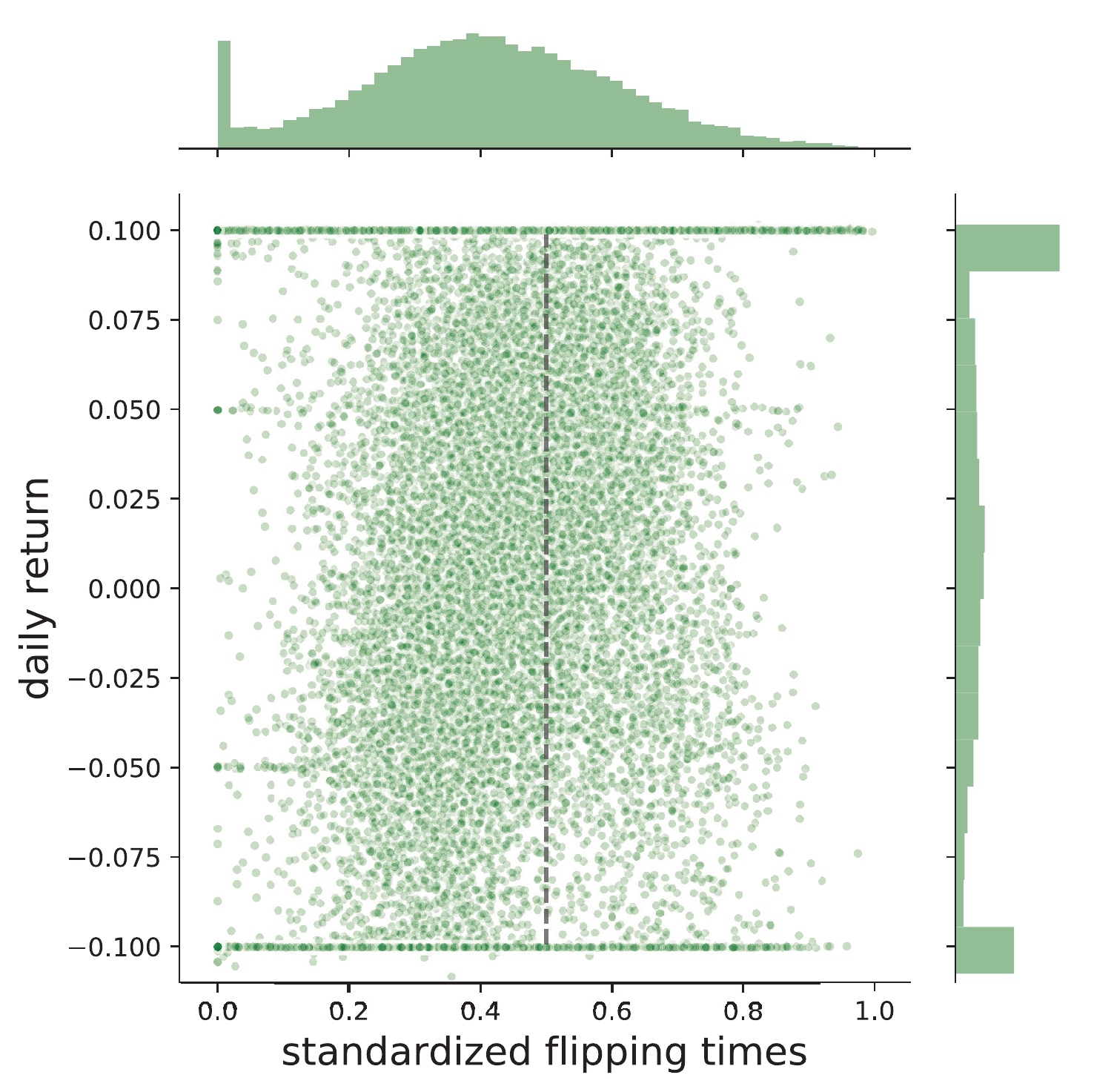}
	\end{minipage}\hfill
\caption{{\bf Polarity flipping times and daily returns.} The distributions on the y-axises are daily returns of all stocks during the specific period. In bull market (a), the returns are almost symmetrically distributed along zero. In the crash period (b), there were lots of stocks dropped to their lower limit, in which their daily percentages are -0.1. In post-crash period (c), with government injecting liquidity, some stocks went up to their upper limits while still some reached their lower limits. Therefore, the daily distribution has extreme values on both ends. The distribution of standardized flipping times is respectively demonstrated on the x-axises. It has change from fairly symmetrical shape (a) to right-skewed (b) and again to roughly symmetrical shape (c). The shifting of skewness in (b) implies the loss of liquidity to some degree.}\label{fig:flipping_times}
\end{figure}

To illustrate the effect of the sign changes of the indicator, we consider returns, incorporating both daily and stock-level measures. The daily return is simply obtained by $(p_{i,d}-p_{i,d-1})/p_{i,d-1}$ for stock $i$ on day $d$. Fig.~\ref{fig:flipping_times} presents the standardized flipping times and daily return of stock $i$ in day $d$. The observations are classified by different market periods. One small thing is that, there are times when the stocks reach their lower limits or upper limits (in China stock market, the allowed maximum one-day drop or rise of a stock is ten percent of its closing price last day), resulting in some samples concentrating around $\pm0.1$ of the y-axis. Apart from this, we find explicit shape of the relationship between flipping times and daily return in pre-crash period, shown in Fig.~\ref{fig:flipping_times}(a). Stocks on the whole exhibit kind of symmetry along the dashed line of 0.5. The stock daily returns are sensitive to flipping times. When it is less than 0.5, the increase in standardized flipping times is associated with an increase in daily return. On the contrary, for standardized flipping times greater than 0.5, the increase of flipping times is associated with a decline in daily return. Hence, rational investors should have a great incentive to sell when they observe that there's a mid-level frequency of polarity flipping in the face of bull market, as the daily returns are high.

The crash period, however, demonstrates no such clear pattern, as can be seen in Fig.~\ref{fig:flipping_times}(b) and (c). Their distribution of flipping times is right-skewed, especially during the first cycle of crash. This indicates that the polarities lean toward one direction when market experience panic. The post-crash period turns out to be a mess and the pattern failed to tell the association.

\subsubsection{Flipping depth}

We now turn to the effect of the polarity flipping amplitudes, called flipping depth. The depth of stock $i$ in day $d$ is defined as

\begin{equation}
\textrm{flipping depth}_{i,d} =\sum_t{|\textrm{polarity}_{i,t,d}-\textrm{polarity}_{i,t-1,d}|},
\end{equation}
where $t$ is the time that polarity flips in day $d$. For comparison reasons, we average flipping depth by flipping times. Thus, we have

\begin{equation}
\textrm{averaged flipping depth}_{i,d} =\frac{\textrm{depth}_{i,d}}{\textrm{number of flipping times}_{i,d}}.
\end{equation}

The averaged flipping depth obtained in this way are caused by cumulative polarity flipping from either selling or buying party to the counterparty one. It reveals the strength of the investors' trading desire shifting.
 
\begin{figure}[h!]
\centering
\includegraphics[width=1\linewidth]{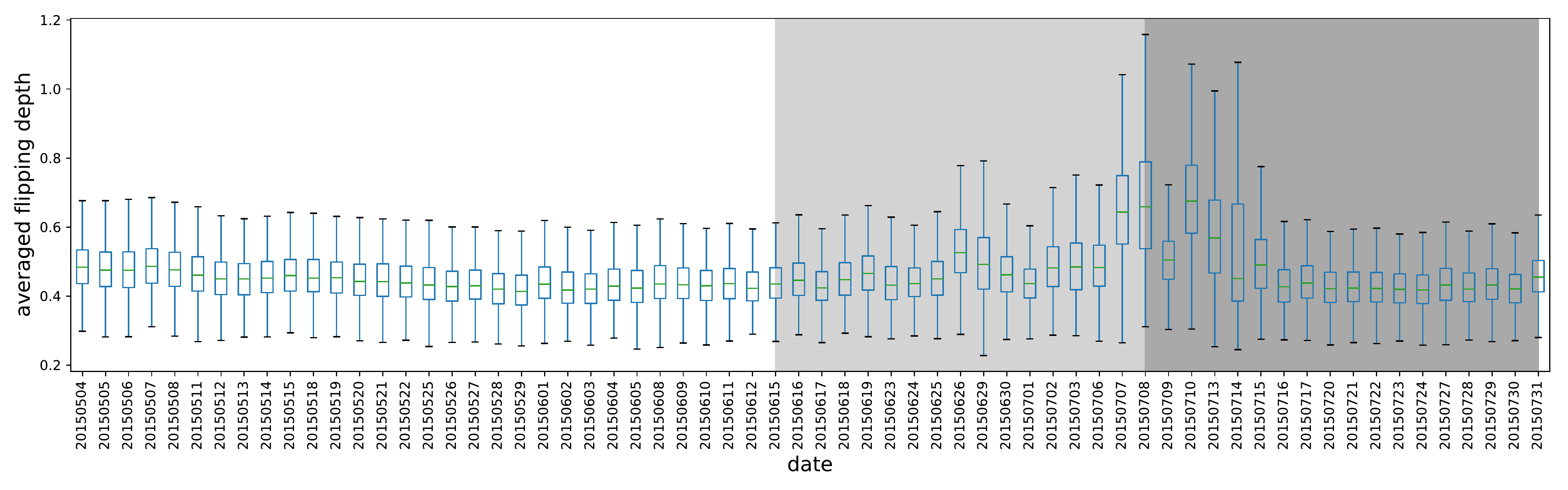}
\caption{{\bf Polarity flipping depth.} The box-plot graphically depicts groups of numerical depths of 1646 stocks in the corresponding day through their five number summaries: the smallest observation, lower quartile ($Q1$, 25th percentile), median ($Q2$, 50th percentile), upper quartile ($Q3$, 75th percentile), and largest observation. Denote the spread between $Q3$ and $Q1$ as $h$, then outliers are defined as those less than $Q1-1.5h$ or greater than $Q3+1.5h$. In each box of this representation, outliers are ignored to make the graph clear. The bottom-most line represents the $Q1-1.5h$ of the sample, and the upper-most vertical line represents the $Q3+1.5h$ of the sample. The bottom of the box represents the $Q1$, and the top of the box represents the $Q3$, with the line inside the box representing the median. The blanking spaces between the different parts of the box help indicate the degree of dispersion and skewness in the data. }
\label{fig:flip_depth}
\end{figure}

Fig.~\ref{fig:flip_depth} gives the whole picture of day-to-day averaged flipping depth distribution. Each box represents the depths of 1646 stocks in the corresponding day. It is obvious that the pre-crash and post-crash periods demonstrate significantly unusual behavior, with the corresponding boxes rising to the upward. The interesting changing of phase behaviors of the market system could be viewed as the outcome of transitions on both investors' expectation and trading preferences, as observed in a complex adaptive system \citep{wang2009heterogeneous}. It can be noticed that the first significant change happened on June 26 2015, in which the stock market had an abrupt sharp fall. It is also apparent that at the beginning of the post-crash period, things worked much differently than before. The amplitude of investors' trading polarity shifting becomes larger, and the variation of this amplitude itself is bigger, indicating there are severe imbalanced in selling and buying activities and the extent of imbalance varies among stocks.

\subsubsection{Length before flipping}

Despite the times and depth of polarity flipping, the duration for the trading polarity to break out of the dominant direction of polarity is of great interest. We call it length before flipping. Suppose we have a five-minute polarity series such as $(0.2, -0.3, -0.4, -0.2, 0.3)$, then the negative polarity length is three, as the number of negative polarity between two positive polarity is three. The length before flipping reflects how long the negative polarity has the domination consecutively. Specifically, a negative length of stock $i$ encompasses two flips is the time span (as we have the polarities equidistantly recorded at one-minute frequency) of negative polarity in between. The same applies to positive polarity length.

\begin{figure}[t!]
	\begin{minipage}{0.5\linewidth}
		\centering
		{\footnotesize (a)}
		\includegraphics[width=1\linewidth]{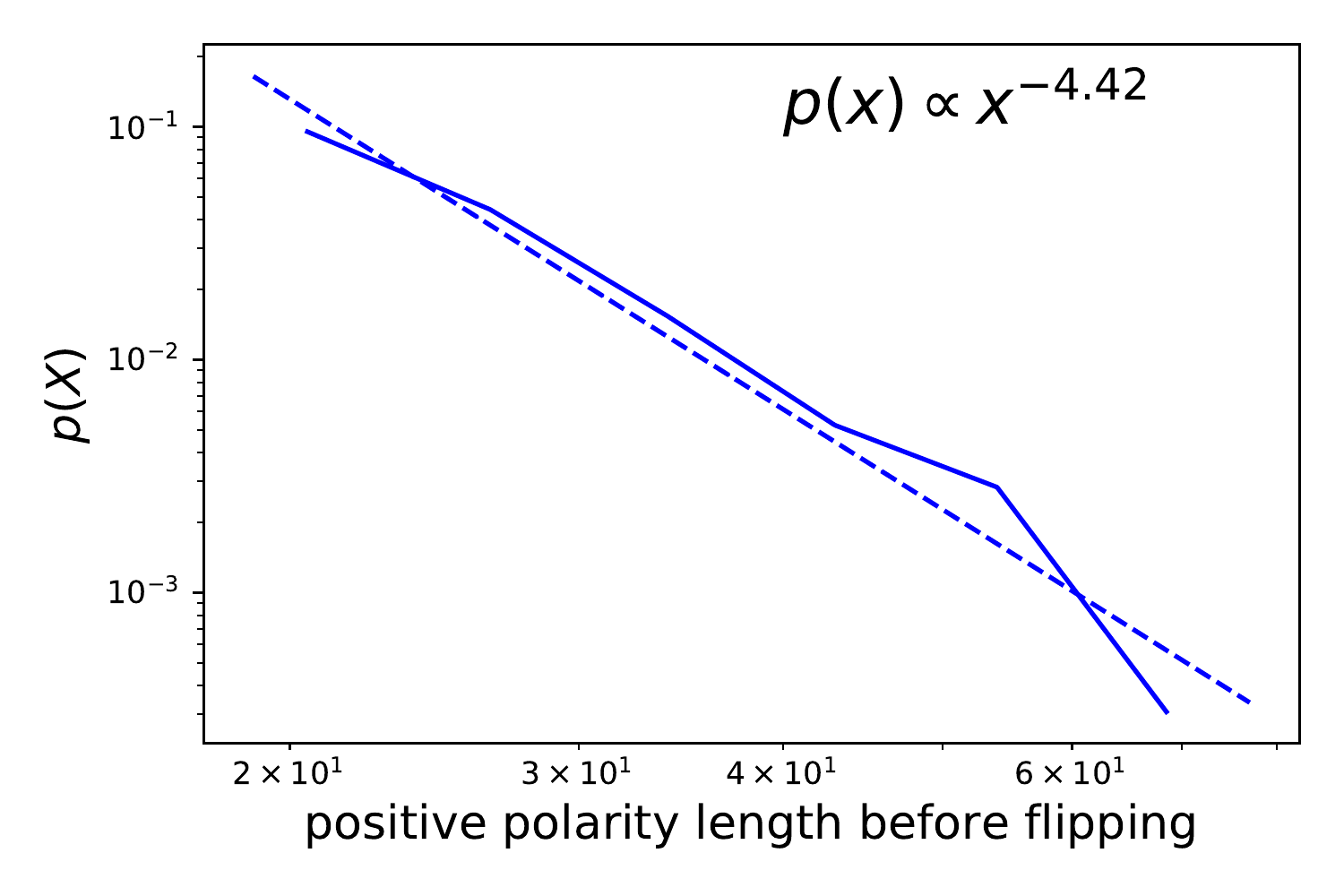}
	\end{minipage}\hfill
	\begin{minipage}{0.5\linewidth}
		\centering
		{\footnotesize (b)}
		\includegraphics[width=1\linewidth]{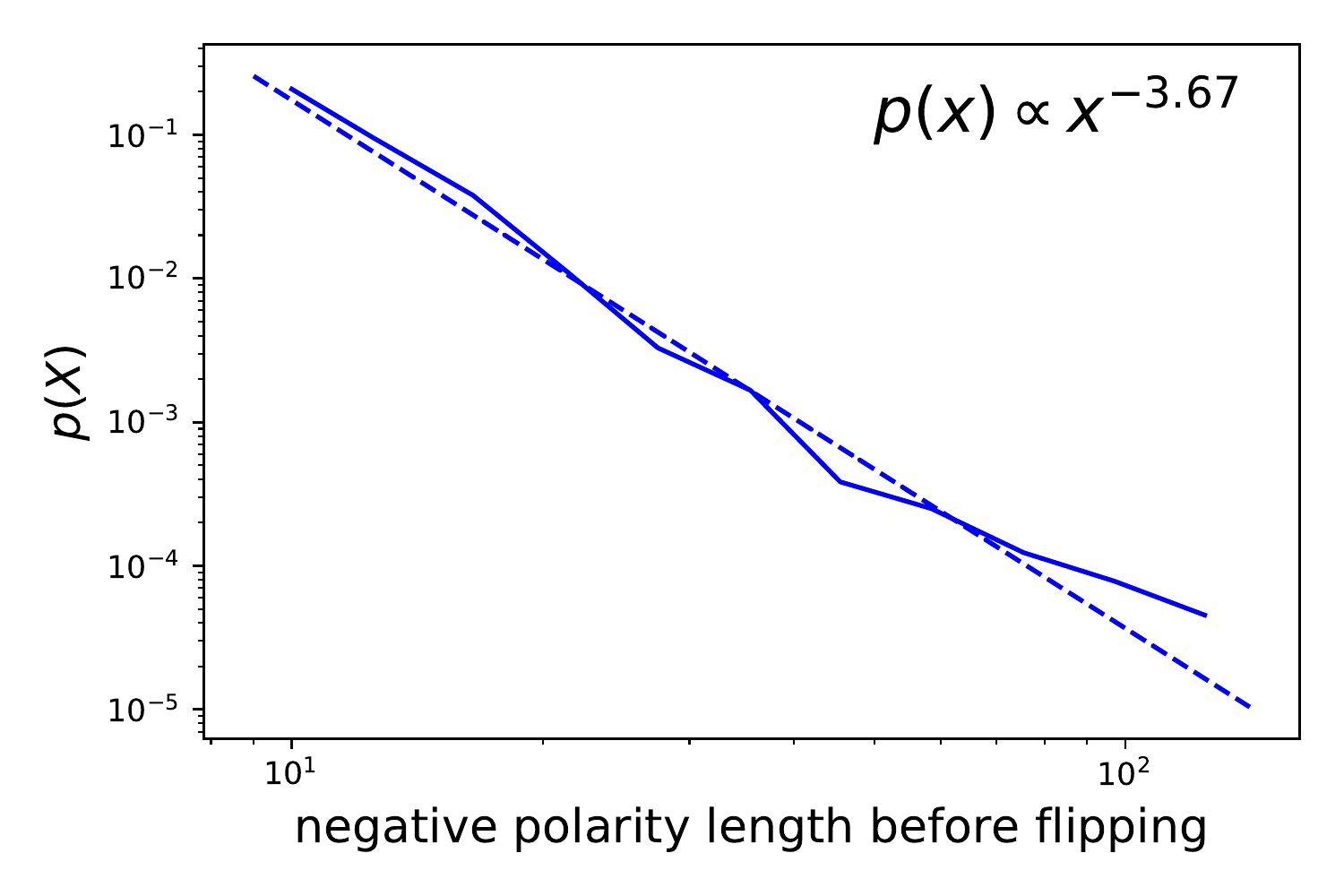}
	\end{minipage}\hfill
\caption{{\bf Probability density function ($p(X)$, blue) of length before flipping and fitted power-law distribution on May 8th 2015.} The subfigure (a) shows the positive flipping length distribution of all stocks, as where the positive flipping length is defined as the time span between two positive polarities. The subfigure (b) shows the negative flipping length distribution of all stocks, as where the negative flipping length is defined as the time span between two negative polarities. The dashed lines are power-law fitting.}\label{fig:508pl}
\end{figure}

\begin{figure}[h!]
\centering
\includegraphics[width=0.95\linewidth]{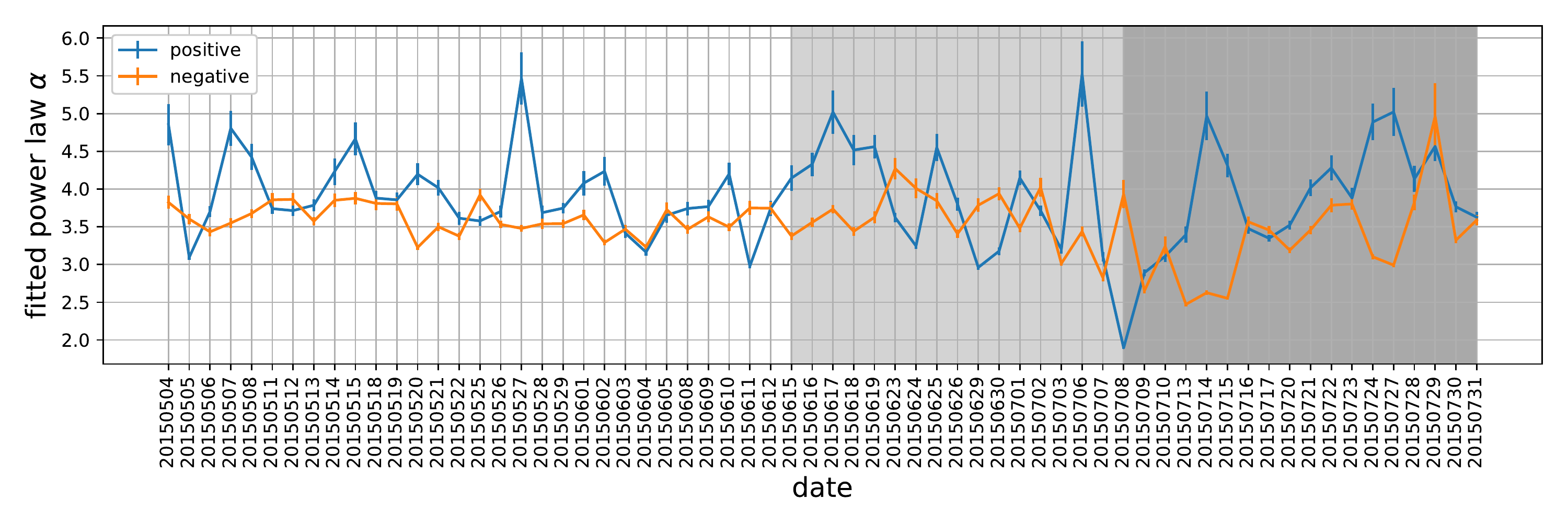}
\caption{{\bf The fitted power-law exponent $\alpha$ of positive and negative flipping length distribution for each day.} The daily distribution is the mixture of all stocks' daily flipping lengths. Error bars are the estimation standard errors for $\alpha$ for the corresponding day.}\label{fig:powerlaw_alpha}
\end{figure}

When mixing all stocks' daily flipping lengths together, we obtain the distribution of flipping length for each day. Fig.~\ref{fig:508pl} is an example. Surprisingly, we find that the distribution is well fitted power-law (the parameter estimation procedure follows \citep{alstott2014powerlaw}), both for positive flipping length and negative flipping length. As we observe similar effects not only for May 8th 2015, but also for other days in our dataset, we conclude that power-law scaling behavior of polarity flipping length is a universal feature. We plot the trend of daily power-law exponent in Fig.~\ref{fig:powerlaw_alpha}. As it can be seen, the negative one changes a lot from the pre-crisis period to crisis period. The positive one, however, have vibrations in both the three period. This shows that the two play completely different roles. Nevertheless, most of the fitting exponents are between 3 and 5, implying a prevalent law of heavy tails in flipping length.

 Notice that, the flipping length in essence can be treated as the time interval between two consecutive flips either from negative to positive or vice versa. Hence the power-law distribution implies that the domination direction of polarity is bursty. Bursts, which have been observed in a wide range of human related systems, indicate the enhanced activity levels over short periods of time followed by long periods of inactivity \citep{barabasi2005origin}. In the present scenario, the observed bursty is the rapid vibrations between the two directions of polarity separated by long periods of one direction's domination. Following \citep{goh2008burstiness}, here we use the burstiness parameter $B=\frac{\sigma_\tau-\mu_\tau}{\sigma_\tau+\mu_\tau}$ to obtain the extent of the bursty for the flipping lengths, where $\mu_\tau$ and $\sigma_\tau$ are the mean and the standard deviation of the daily power-law distributions. The magnitude of the parameter correlates with the signal's burstiness. When $B = 1$, the series is the most bursty signal. When $B = 0$, the series is neutral, and when $B = -1$ corresponds to a completely regular (periodic) signal. 

\begin{figure}[h!]
\centering
\includegraphics[width=0.95\linewidth]{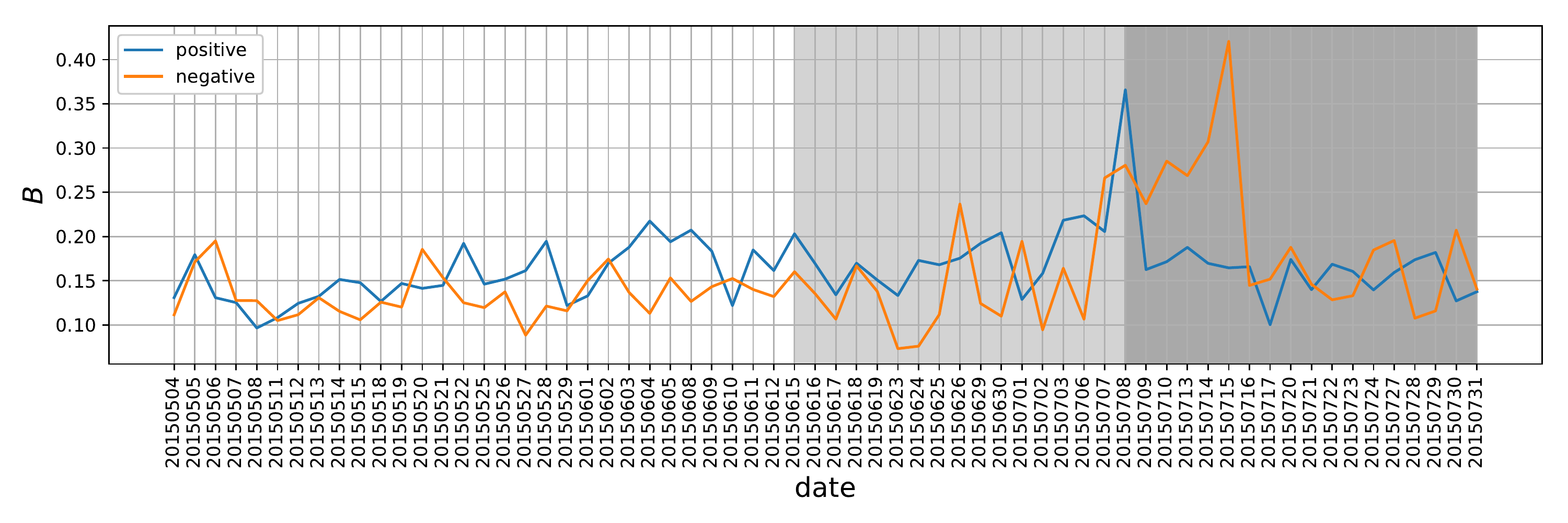}
\caption{{\bf The burstiness parameter $B$ for the daily distributions of positive and negative flipping length.} The daily distribution is the mixture of all stocks' daily flipping lengths.}\label{fig:bursty}
\end{figure}

The trends of burstiness can be found in Fig.~\ref{fig:bursty}. In the pre-crash period, both the positive and negative flipping length distributions have roughly stable performance. The negative flipping length occurred the first burstiness hop on June 26, when the market experiencing the most severe crash. The higher value of burstiness revealed by the hop implies that the selling polarity of stocks prevailed longer in the trading hours, which is consistent with the enduring panic selling happened on that day. Meanwhile, at the beginning of post-crash, the burstiness of negative flipping length became larger than before, indicating the market sell-off was getting more severe.

Interestingly, burstiness of positive and negative flipping length get to a high level at the same time on July 8th, when China State-owned Assets	Supervision and Administration Commission prohibited state-owned companies reducing stocks holding-shares and China Securities Regulatory Commission declared to increase the purchasing of mid-cap and small-cap stocks. It conveys that the powerful bailout measures announced on that day had led the trading behavior to be poles apart among stocks: whether the selling man-times dominate the trading or buying man-times dominate the trading. This probably due to the co-existing of two polarized expectations in face of government interventions, of which some people were encouraged by the `national team' and decided to buy while other decided to take the opportunity to sell their holdings. However, the burstiness of positive flipping length dropping sharply on the following day demonstrates that the market confidence did not last long than expected.

It seems that burstiness parameters are more sensitive than the power-law exponents in the observed system, as significant differences are exhibited in pre-crash, crash, and post-crash periods. What's more, the observed bursty character reflects some fundamental and potentially generic feature of market participants' trading dynamics at the times of market interventions. This again demonstrates that it is feasible to develop a signal that could systematically reflect trading psychology and emotion, through the proposed trading polarity that originates from micro-level data. 

Overall, the discussions on polarity flipping behavior have shed lights on trading polarities' essence nature of micro trading behavior. In the next section, we move on to its interconnections with investors' emotion and stocks' returns from the prospectives of market-level as well as stock-level.

\section{Market-level polarity}\label{subsec:marketlevel}

\subsection{Market emotion and polarity}

\begin{figure}[h]
\centering
\includegraphics[width=0.6\linewidth]{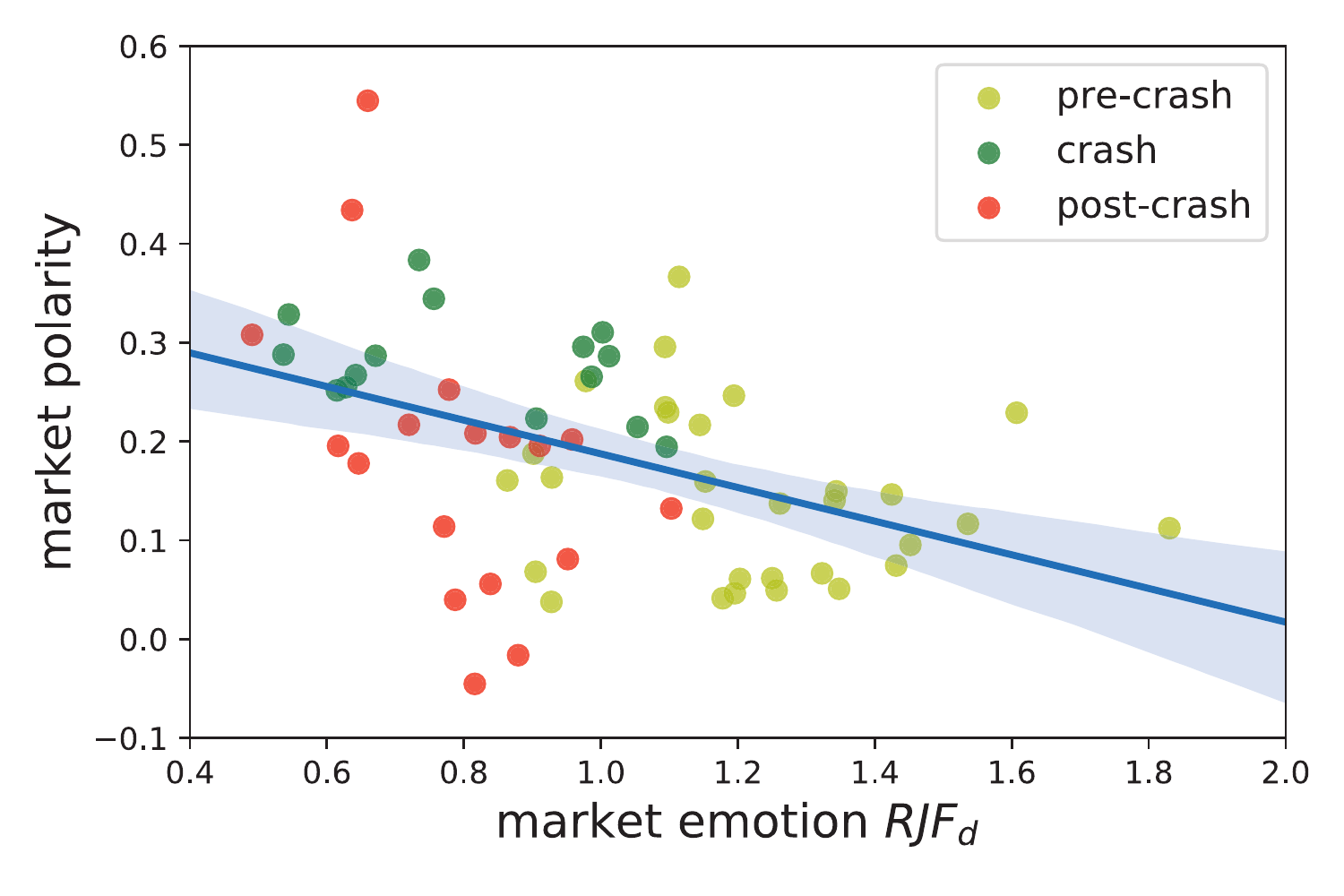}
\caption{{\bf Trading polarity value correlated with market emotion indicator.} Based on online emotions of investors, $RJF_d$ on the x-axis is defined as the ratio of joy (greed) to fear ($RJF$) in day $d$ as $RJF_d=\frac{X_{joy,d}}{X_{fear,d}}$. The market polarity on the y-axis is the averaging polarity of stocks when the SZSC index reaches its daily lowest.}\label{fig:emotion}
\end{figure}

From a systematic view, this section investigates the polarity's interconnection with emotion. The daily market emotion of China stock market is measured by $RJF_d$ \citep{zhou2017}. Based on online emotions of investors, $RJF_d$ is defined as the ratio of joy to fear in day $d$ as $RJF_d=\frac{X_{joy,d}}{X_{fear,d}}$. While $RJF_d$ is greater than 1, the investors are optimistic and consider the market is rising. On the contrary, while $RJF_d<1$, investors are irrational, fear is the dominant emotion in the market and investors are afraid of the loss of benefit. The investors stay rational and the emotions are stable as $RJF_d$ is around 1. The daily $RJF_d$ is shown on x-axis in Fig.~\ref{fig:emotion}.

The polarities of stocks are averaged to obtain the market-level polarity at the specific time. That is, the average of these polarities across the stocks in the sample as the polarity on the equal-weighted market portfolio to reflect the imbalance at the level of market, i.e.,

\begin{equation}
\textrm{market polarity}_{[t-1,t],d} =\frac{1}{N} \sum_{i=1}^{N} \textrm{polarity}_{i,[t-1,t],d},
\end{equation}
where $N$ is the number of stocks in the dataset. We consider SZSC index to represent the market trend. From the \emph{intraday} one-minute index return series, we pick out the minimum return in each day, and use the market polarity at that same specific moment to represent the daily polarity. The reason it that, as \citep{zhou2017} has pointed out, the online emotion is more sensitive to bad market state. Therefore, the polarity when market reaches the lowest point could serve as the daily polarity to get along with daily market emotion. The polarities are exhibited on the y-axis in Fig.~\ref{fig:emotion}.

Fig.~\ref{fig:emotion} presents the scatterplot between emotion and polarity, along with the line of best fit. We find that most of the market polarity is above zero, implying that the buying polarity is the dominant role when market reaches its worst point of each day. More importantly, the figure illustrates the negative correlation between emotion indicator and market polarity. That is, as the market getting less optimistic, the polarity moves away from the balanced states to buying polarity, suggesting a generic buying opinions imbedding into the market.

To get a closer look, the yellow dots are those in pre-crisis period, when the market kept going up whereas market emotion is excited in most of the time. The green and red dots that from crash and post-crash periods are depart from the fitted line. If we separately compute the correlation coefficients for the three period, we get -0.197, -0.382, -0.493, respectly. Though the number of dots in every period may not be enough to get a linear fit, we could still roughly find that the correlation pattern turns sharper in the crisis period and in the post-crisis period.

It is worth claiming that, accumulating enough expressions, either texts in social media \citep{zhou2017,xu2017weibo} or quires in search engine \citep{bordino2012web}, is always necessary for direct measurement of the market sentiment. However, in the circumstance of high-frequency analysis, it is unfortunately challenging for the sentiment analysis due to the severe sparsity of emotional expressions in such a short interval. While the significant correlation disclosed here implies that our presented polarity, which essentially calculated in the context of high-frequency trading, can break the above constrains and real-timely reflect the sentiment of the market.

\subsection{Market return and polarity}

We further investigate the relationship between polarities and returns. Still, we regard the SZSC Index as the market-level return. For the purpose of calculating the percentage change of the index, we use the last price of every one minute from the \emph{intraday} data. The value of index at time $t$ on day $d$ is $p_{t,d}$. We use the last price of a day from the \emph{end-of-day} data as the baseline price for computing the return next day, denoted as $p_{d}$. The percentage change of index at time $t$ on day $d$ is computed by $(p_{t,d}-p_{d-1})/p_{d-1}$. The reason is that this kind of percentage changes is consistent with what investors see during trading hours on any trading information board, which could stir up tensions and impact the trading behavior directly. 

\begin{figure}[h]
\centering
\includegraphics[width=0.6\linewidth]{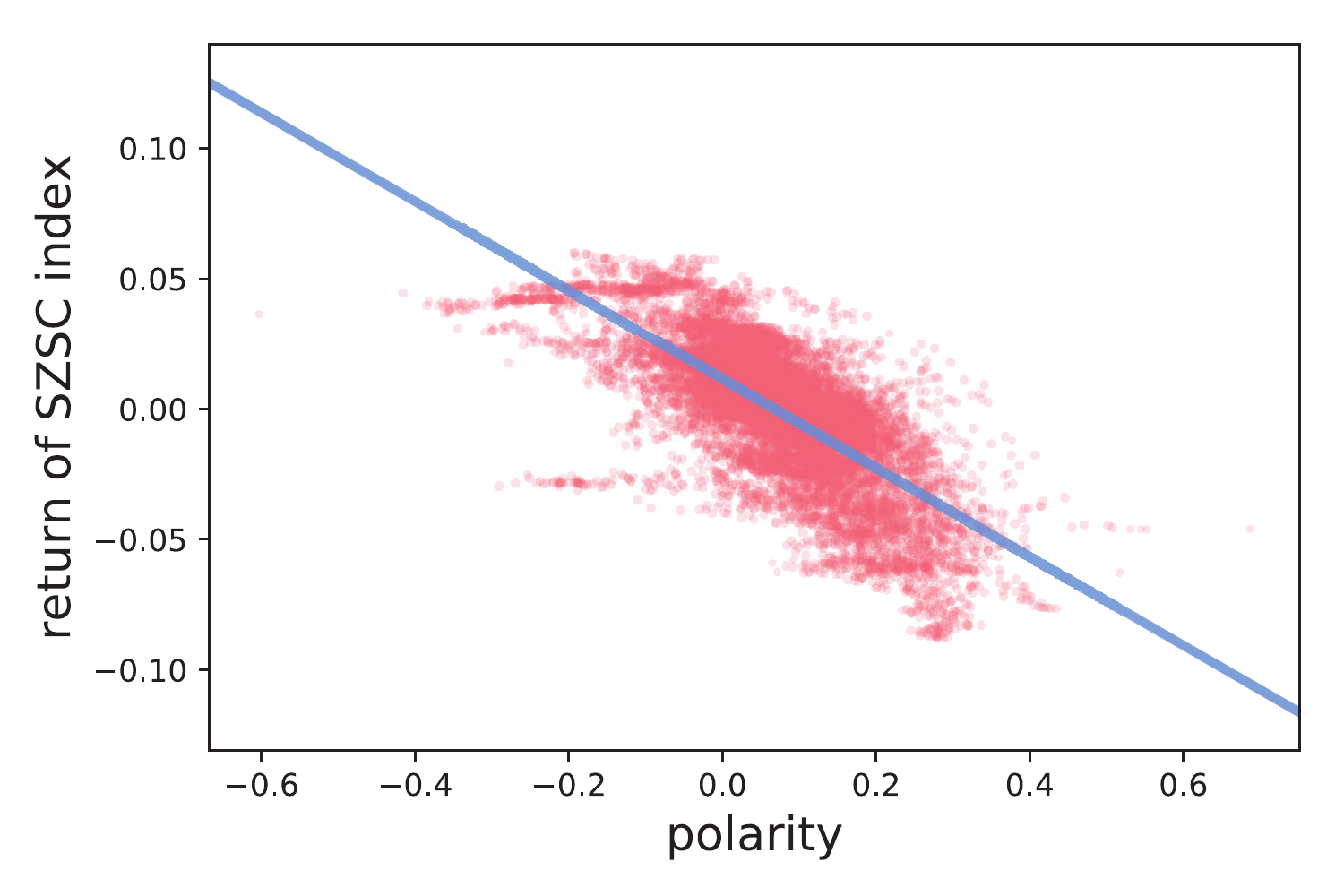}
\caption{{\bf Market-level polarity and return of market index.} The percentage change of index at time $t$ on day $d$ is computed by $(p_{t,d}-p_{d-1})/p_{d-1}$. Market-level polarity is the averaged polarity of all stocks in Shenzhen Stock Exchange. For the two market-level indicators, we use the one-minute frequency and they match each other minute-by-minute.}\label{fig:market_scatter}
\end{figure}

The relationship between market-level polarity and return of market index is shown in Fig.~\ref{fig:market_scatter}. Each dot represents the values on the two dimensions at one specific minute. Given the sampling period of 64 trading days and 4 trading hours per day, there are over 15000 dots in the plot. As can be seen the market return is negatively correlated with market polarity. The Pearson correlation coefficient is -0.72 in this case. We can already see that the negative correlation are clearly present even at an exceedingly general level of analysis.

It is excepted that the market polarity would have influence on the market return. But unfortunately, through Granger causality test \citep{granger1969investigating,toda1995statistical}, by using the return based on neither the last price of yesterday or the last price of the former minute, the index return series only have 0.25 (using the return that calculated based on the last price of the former minute) or 0.28 (using the return that calculated based on the closing price of yesterday) pass rate. Note that the pass rate indicates the proportion of days that market polarity Granger-causes SZSC index return among the sampling days. On the contrary, the polarity Granger-causes return hypothesis has pass rate of 0.97 (using the return that calculated based on the last price of the former minute) and 0.95 (using the return that calculated based on the closing price of yesterday). This means that the polarity may not be capable to predict the return, but it is influenced by the return. It suggests that the market-level trading behavior does respond to the market return. Note that the Granger causality test could not be extended to stock-level analysis, as some of the stocks have no trades at all in some trading minutes and with these missing values it is not possible to apply Granger causality test using the same procedure. Nevertheless, the strong negative correlation between market polarity and market return and the interplays implied by Granger causality test both provide evidence that the polarity indicator can capture the market behavior that are bound up with market return. It serves as an effective behavioral signal at market-level.

\section{Stock-level polarity}\label{subsec:stocklevel}
\subsection{Immediate price impact}\label{subsec:price_impact}

For a given trading polarity within one minute of one stock, the price impact could be obtained by its return within the minute. We calculate the return for each stock in every one minute, making it match the polarity. Denote the price of stock $i$ at time $t$ in day $d$ as $r_{i,t,d}$, where $r_{i,t,d}=log(p_{i,t,d})-log(p_{i,t-1,d})$, which commonly used in most financial studies. For every $r_{i,t,d}$, there's one and the only polarity of $polarity_{i,t,d}$ accordingly. These pairs of polarity and return are sorted into three groups according to the sign of polarity, regardless of the stocks and time: the zero group includes pairs of polarity and return with polarity equal to zero; the negative group contains the polarity value less than zero, and the positive group contains the polarity value greater than zero. The reason is that, the three kinds of polarity imply complete different directions of trading concentration and trading interest. 

\begin{figure}[h]
\centering
\includegraphics[width=0.6\linewidth]{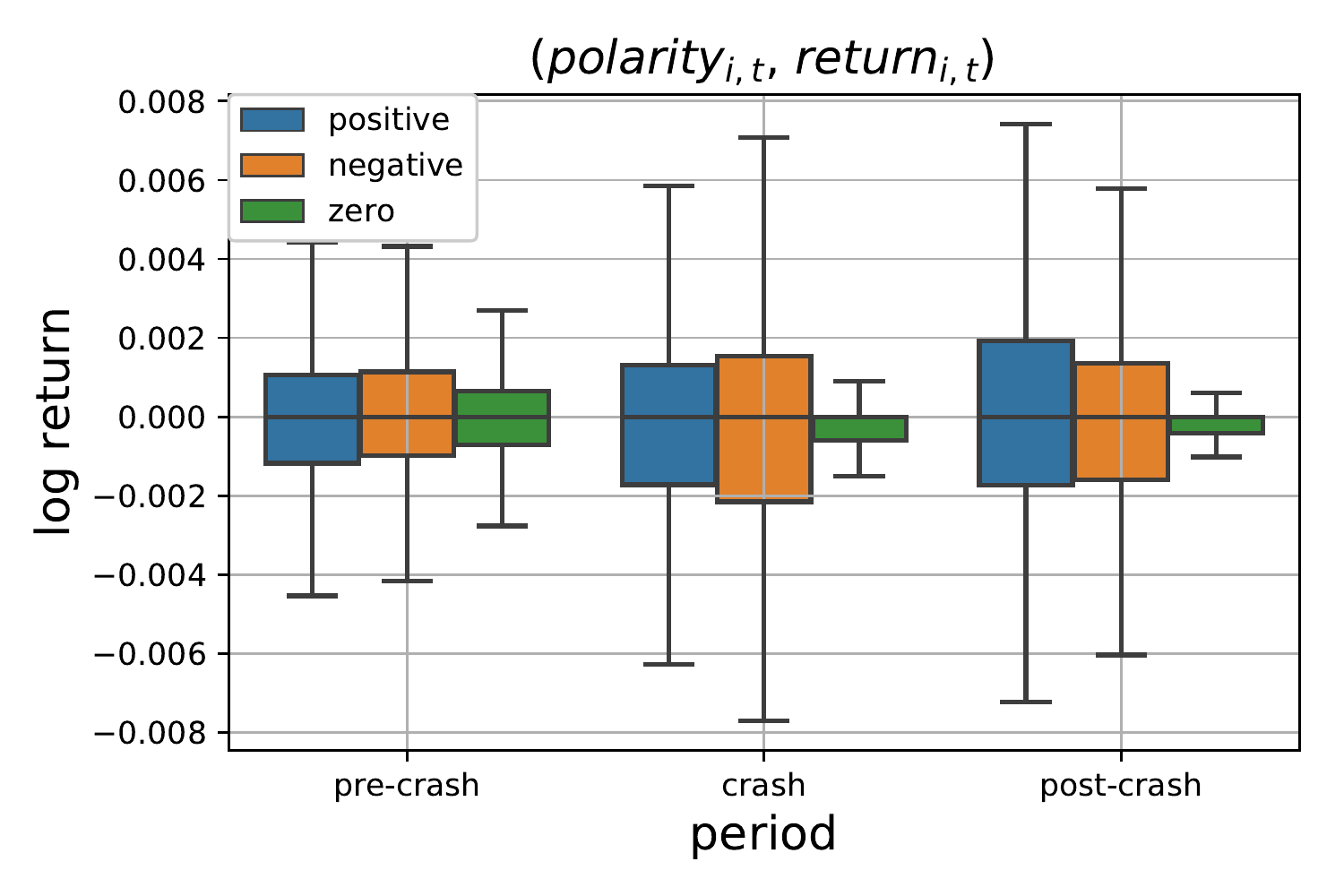}
\caption{{\bf Immediate price impact.} The pairs of polarity and return on a one-minute basis are sorted into three groups according to the sign of polarity. The box-plot graphically depicts groups of numerical log-returns through their five number summaries: the smallest observation, lower quartile ($Q1$, 25th percentile), median ($Q2$, 50th percentile), upper quartile ($Q3$, 75th percentile), and largest observation. The spacings between the different parts of the box help indicate the degree of dispersion and skewness in the data. Denote the spread between $Q3$ and $Q1$ as $h$, then outliers are defined as those less than $Q1-1.5h$ or greater than $Q3+1.5h$. In each box of this representation, outliers are ignored to make the graph clear. The bottom-most line represents the $Q1-1.5h$ of the sample, and the upper-most vertical line represents the $Q3+1.5h$ of the sample. The bottom of the box represents the $Q1$, and the top of the box represents the $Q3$, with the line inside the box representing the median. }\label{fig:price_impact}
\end{figure}

In Fig.~\ref{fig:price_impact}, we use box-plots because it is possible to present both the median and the entire spread of the sample population for different groups. As can be seen, whatever groups the polarity belongs to, returns are centered around zero. Comparing to positive and negative polarity groups, the zero-polarity is less dispersed, indicating the balanced relationship between buying and selling is stabilizing stocks prices. For the positive and negative polarity groups, results are mixed temporally. In bull market, positive polarity group is slightly lower than the negative group, as where the dispersions of the two are nearly the same. Intuitively, this indicates that the selling man-times overwhelming buying man-times is more likely to bring about higher returns. On the contrary, the two crash periods both have much more variations in positive and negative polarity groups. These imply that the polarizations inspire more uncertainty in market crash. Furthermore, the negative group is more dispersed than positive group during the first round of market crash. Moving towards the post-crash period, however, things are just the opposite, where the positive group is more dispersed, implying the polarizing toward buying has exerted much variations in stock returns in the second round of market crash.

\subsection{The changing of polarity-return correlation}
\begin{figure}[h]
\centering
\includegraphics[width=0.8\linewidth]{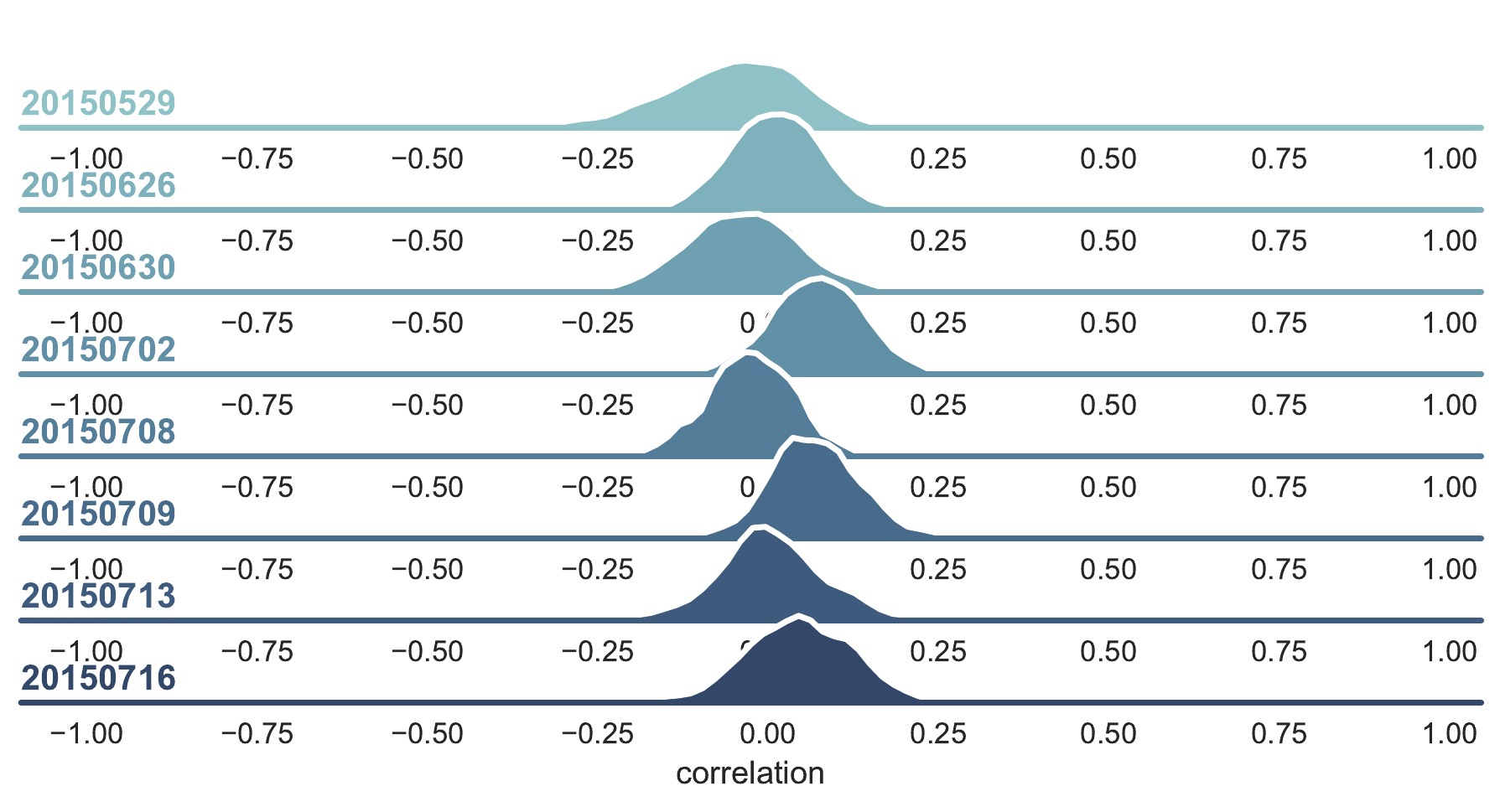}
\caption{{\bf Correlation distribution in a couple of days.} Using the one-minute frequency data, we get the whole set of polarity-return correlation coefficients for all stocks in day $d$. The distributions of correlations in different days exhibit in rows, respectly. }\label{fig:example}
\end{figure}

By constructing the Pearson correlation coefficient of polarities and log-returns of each stock on each day, using the one-minute frequency data, we get the whole set of correlation measures for all stocks in day $d$, and its probability distribution is denoted as $Q_d(x)$. Fig.~\ref{fig:example} shows examples of a few days of correlation distributions, which basically demonstrates the correlation do vary from time to time, as where some of them are left skewness, but others are right skewness, and some are fat tailed while others are not. As trading polarity origins from market microstructure, we argue that the microstructure would have changed under various market conditions. 

To access the disparity of correlation distribution from day to day, we construct Kullback-Leibler divergence measures on daily correlation coefficients distribution. Denote the probability distributions in day $d$ and $d-1$ as $Q_d(x)$ and $Q_{d-1}(x)$. The Kullback-Leibler divergence is measured by to be from $Q_{d-1}$ to $Q_d$ is defined by

\begin{equation}
KL(Q_d||Q_{d-1})=\sum_x{Q_d(x)\log\frac{Q_d(x)}{Q_{d-1}(x)}}.
\end{equation}

\begin{figure}[h]
\centering
\includegraphics[width=1\linewidth]{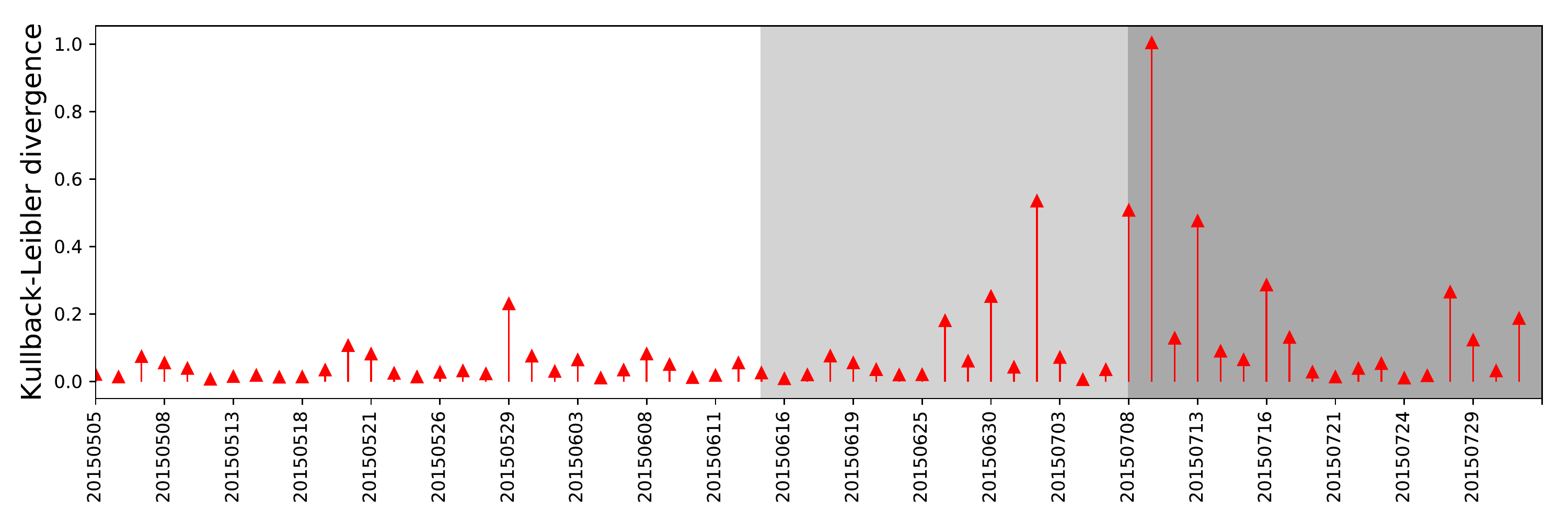}
\caption{{\bf The Kullback-Leibler(KL) divergence from $Q_{d-1}(x)$ to $Q_d(x)$.} The higher the KL divergence, the more diverse the correlation distribution is comparing to yesterday. It is clearly shown that the KL divergence increases in crash and post-crash period.}\label{fig:kl}
\end{figure}

The divergence of correlation distributions is shown in Fig.~\ref{fig:kl}. The shaded areas correspond to the stock market crisis in June to July 2015. It clearly shows that the KL divergence increases in bad times, widening the difference from the day before that. The first notable difference is June 26, when the market suffer from over one thousand of stocks falling to their lower limits. The most significant changes happened around the beginning of post-crash times, when government started to take measures to save the market. From this point of view, the measure is efficient in telling the phase transitions in the underlying correlation between trading behavior and stocks returns. It could be used to signal the changing of market.

\section{Conclusion}\label{subsec:conclusion}

With the rapid growth of financial data and the magnificent development in computational ability, studies driven by big-data and dense-computing in recent decades indeed offer better understandings of the financial market \cite{wood1985investigation, dufour2000time,ivanov2014impact,xie2016quantifying,lillo2003econophysics,preis2011switching,bhattacharya2017superstitious,menkveld2017shades}. Rather than the heavy depending on theoretical assumptions, data-driven solutions can model the market from a more systematic and realistic view, making the real-time and precise reflection or even prediction of financial systems possible in real-world scenarios \cite{wood1985investigation,huang2003trading,kenett2010dominating,ivanov2014impact,xie2016quantifying,preis2011switching}. 

Under these circumstances, we propose an indicator, called trading polarity, to depict the imbalanced relationship between buying and selling in the granularity of man-times. From the initial investigations, we find that it differed by stocks capitalization. Furthermore, we have assemblies of indictors that tell us about the trading polarity, including polarity ratios, flipping times, flipping depth, flipping length, and the distribution of correlation between stock's polarity and return. We have shown that these measures could bear a more meaningful relation to the changing of market conditions, especially during extreme market crash. The usefulness of this framework is that it deepens our understanding of trading patterns, which originates from `micro' data but surprisingly possesses the explaining power at `macro'-level. Also, as discovering the vibratory rate of a market gives one the key to trade it efficiently. 

As for the predictive ability of stock's polarities for returns, our results could not give sufficient evidence to answer the question of whether polarity in stocks have significant implications on returns. We do, however, perform one analysis that goes beyond the stock-level and examines the polarity-return relation in the aggregated market-level. Interestingly, we find that the market-level polarity does not lead the return of market index, but responds to it. This provides evidence that the polarity indicator can capture the market behavior that are bound up with market return. More importantly, we find the convincing correlation between market polarity and market emotion. This again implies that the proposed trading polarity, that could easily be calculated in the context of high-frequency data, can provide the measurement of sentiment for the market in a real-timely way. In a word, it could serves as an effective behavioral signal at market-level.

Being such an indicative indicator of macro trading pattern, the proposed polarity could be of interest to market regulators. Broadly speaking, regulators should be aware of polarity in the stock market. Firstly, attempts are needed to regulate imbalanced polarity that are likely to push the price away from fundamental prices, or potentially increase the crash risk. Secondly, regulators should gain knowledge of polarity pattern in stock trading as a whole and in cases where there are externalities from traders to market prices and real economic activity. These issues call for more research with longer-term data in the future. What's more, how to integrate volume into the current analysis is also meaningful given that trading activity has usually been proxied by volume \citep{lillo2003econophysics}, and further research should consider it in the presented framework. Also, if data is available, we could compare the difference characteristic of markets in different countries. Nevertheless, the present study could be of interest for both policy makers and market participants.

\section{Acknowledgments}
This research was financially supported by National Natural Science Foundation of China (Grant Nos. 71420107025 and 71501005). The authors also appreciate the constructive comments from Mr. Zhenkun Zhou.



\end{document}